\documentclass[12pt]{article}

\usepackage{graphicx}   
\usepackage{amsmath} % assumes amsmath package installed
\usepackage{amssymb}  % assumes amsmath package installed
\usepackage[left=2.5cm,top=1.5cm,right=2.5cm,bottom=3cm]{geometry}
\usepackage{multirow}
\usepackage{amsmath} % assumes amsmath package installed
\usepackage{amssymb}  % assumes amsmath package installed
\usepackage{amsthm}
\usepackage{mathrsfs}
\usepackage{algorithmicx}
\usepackage{algpseudocode}
\usepackage[Algorithm,ruled]{algorithm}

\title{\LARGE \textbf{Gaussian Process Regression for Generalized Frequency Response Function Estimation}}
\date{\vspace{-5ex}}

\author{Jeremy Stoddard$^{a}$, Georgios Birpoutsoukis$^{b}$ \\
$^{a}$ University of Newcastle, Callaghan, NSW, Australia \\
$^{b}$ Vrije Universiteit Brussel, Brussels, Belgium}
\date{\today}

\theoremstyle{definition}
\newtheorem{assumption}{Assumption}

\theoremstyle{definition}
\newtheorem{remark}{Remark}

\theoremstyle{definition}
\newtheorem{thm}{Theorem}

\theoremstyle{definition}
\newtheorem{myDef}{Definition}

\theoremstyle{definition}
\newtheorem*{prf*}{Proof}

\begin{document}

\maketitle

\section{Introduction}
\label{sec:GFRF_intro}

Kernel-based modeling of dynamic systems has garnered a significant amount of attention in the system identification literature since its introduction to the field in \cite{pillonetto2010}. While the method was originally applied to linear impulse response estimation in the time domain, the concepts have since been extended to the frequency domain for estimation of frequency response functions (FRFs) \cite{lataire2016}, as well as to the Volterra series in \cite{birpoutsoukis2017regularized}. In the latter case, smoothness and stable decay was imposed along the hypersurfaces of the multidimensional impulse responses (referred to as `Volterra kernels' in the sequel), allowing lower variance estimates than could be obtained in a simple least squares framework.

The Volterra series can also be expressed in a frequency domain context, however there are several competing representations which all possess some unique advantages \cite{cheng2017volterra}. Perhaps the most natural representation is the generalized frequency response function (GFRF), which is defined as the multidimensional Fourier transform of the corresponding Volterra kernel in the time-domain series.  The representation leads to a series of frequency domain functions with increasing dimension.

\section{Review of GFRFs}

\subsection{The Volterra Series in the Frequency Domain}

We consider the nonlinear systems whose output is described by the discrete time Volterra series given by \cite{schetzen1980} ,
\begin{equation}
\label{VolterraTD}
y^0(k) = \sum_{m=1}^{M} \Bigg[ \sum_{\tau_1=0}^{n_m - 1} \hdots \sum_{\tau_m=0}^{n_m-1} h_m(\tau_1,\hdots,\tau_m) \prod_{\tau = \tau_1}^{\tau_m} u(k-\tau) \Bigg],
\end{equation}  
where $h_m(\tau_1,\hdots,\tau_m)$ is the $m$'th order Volterra kernel, $n_m$ is the memory length of $h_m$, and $\tau_j$ is the $j$'th lag variable for the kernel. The subscript $m$ gives the dimension of the kernels, up to a maximum degree, $M$. 

\begin{figure}[h]
\centering
\includegraphics[width=0.9\textwidth]{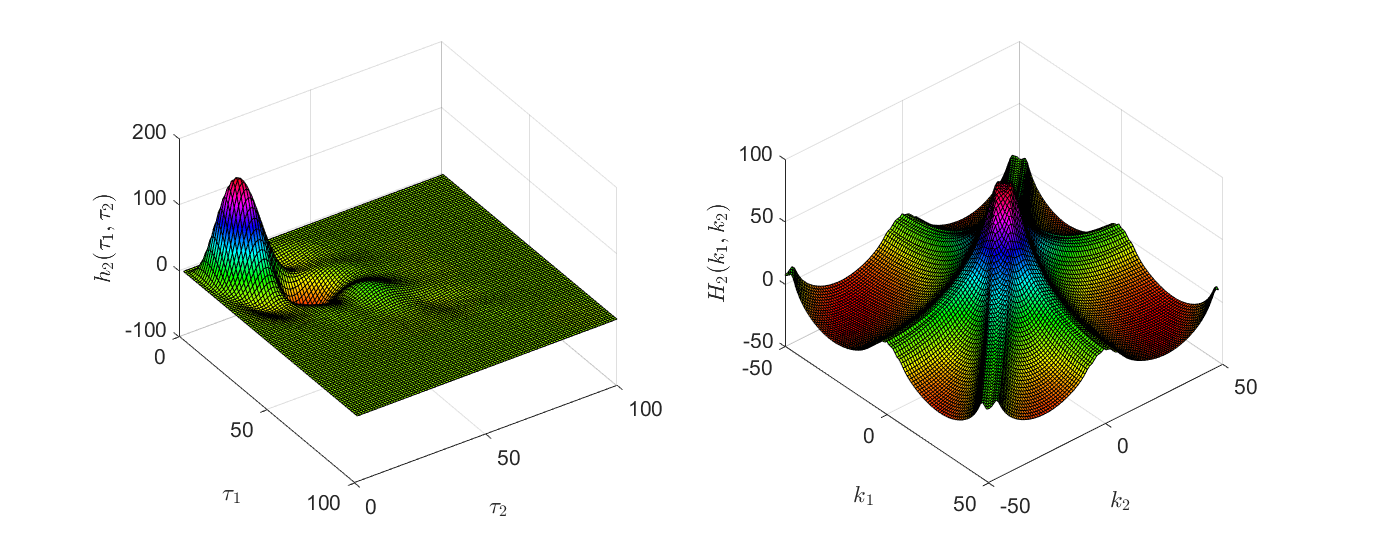}
\caption{Time domain kernel (left) and frequency domain GFRF (right) for a second order Wiener-Hammerstein system}
\label{fig:Kernel_vs_GFRF}
\end{figure}

A steady-state expression for such Volterra series models in the frequency domain was first derived in \cite{george1959continuous}. For $N$-point input and output DFT spectra given by $Y(k)$ and $U(k)$ respectively, the relationship between the two is described by,
\begin{align}
Y^0(k) &= \sum_{m=1}^{M} Y_m(k) \; \; \forall k, \\
Y_m(k) &= \sum_{k_1 + \hdots + k_m = k} H_m(\Omega_{k_1}, \hdots,\Omega_{k_m}) \prod_{i=1}^{m} U(k_i),
\end{align}
where $\Omega_{k}$ is a generic frequency variable corresponding to the $k$'th DFT index, and $H_m(\Omega_{k_1}, \hdots,\Omega_{k_m})$ is labelled the $m$'th order GFRF, given by a multidimensional DFT of the $m$'th time domain kernel, $h_m$, i.e.
\begin{equation}
H_m(\Omega_{k_1}, \hdots,\Omega_{k_m}) = \sum_{\tau_1=0}^{n_m - 1} \hdots \sum_{\tau_m=0}^{n_m-1} h_m(\tau_1,\hdots,\tau_m) e^{\frac{-j2 \pi k_1 \tau_1}{N}} \hdots e^{\frac{-j2 \pi k_m \tau_m}{N}}.
\end{equation}
A second order example is provided in Figure \ref{fig:Kernel_vs_GFRF} to compare the two kernel domains. Direct frequency domain estimation of GFRFs is typically very difficult in practice, due to their multidimensional nature and the complexity of the hyperplane sum. 

%%%%%%%%%%%%%%%%%%%%%%%%%%%%%%%%%%%%%%%%%%%%%%%%%%%%%%%%%%%%%%%%%%%%%%%%%%%%%%%%%%%%%%%%%%%%%%%%%%%%%%%%%%%%%%%%%%%%%%%%%%%%%%%%%%%%%%%%%%%%%%%%%%%%%%%%%%%%%%%%%%%%%%%%%%%%%%%%%%%%%%%%%%%%%%%%%%%%%%%%%%%%%%%%%%%%%%%%%%%%%%%%%%%%%%%%%%%%%%%%%%%%%%%%%%%%%%%%%%%%%%%%%%%%%%%%%%%%%%%%%%%%%%%%%%%%%%%%%%%%%%%%%%%%%%%%%%%%%
\section{Regularized Estimation of GFRFs}

The concepts introduced in \cite{lataire2016} can be extended to estimation of GFRF models, however the covariance functions will differ significantly from the linear case, and the model must now be constructed using a matrix multiplication of the regressor and parameter vector, rather than a simple Hadamard product. This section will describe the Gaussian assumptions on the system, detail the covariance function design and tuning, and derive the maximum a posteriori (MAP) estimates of the GFRFs.

\subsection{Complex Normal Distributions}

In this paper, we will neglect all DFT indices associated with strictly real spectrum quantities, such that the complex normal distribution is sufficient to describe all Gaussian processes in the system. Relevant notation for such distributions is detailed here.

For a complex Gaussian vector, $X$, we define the augmented vector as $\widetilde{X}^T = [X^T X^H]$, where $(\cdot)^T$ and $(\cdot)^H$ denote the transpose and Hermitian transpose of a vector respectively. The distribution of $X$ is written as,
$$ X \sim \mathcal{CN}(m,\Sigma)$$
where $m = \textbf{E}\{\widetilde{X} \}$ and $\Sigma = \textbf{E}\{(\widetilde{X} - m)(\widetilde{X} - m)^H\}$. The augmented covariance, $\Sigma$, can be decomposed into covariance and relation function components, $K$ and $C$, as
$$ \Sigma = \begin{bmatrix} K & C \\ C^H & \overline{K} \end{bmatrix}$$
where $K = \textbf{E}\{ (X - \textbf{E}\{X\}) (X - \textbf{E}\{X\})^H  \}$ and $C = \textbf{E}\{ (X - \textbf{E}\{X\}) (X - \textbf{E}\{X\})^T  \}$.

\subsection{System Assumptions}

For convenient estimation, some assumptions must be made on the system, its Gaussian distributions, and the input excitation.

\begin{assumption}
The (discrete) frequency domain description of the system in steady-state can be given as,
\begin{equation}
\begin{split}
\label{ModelAssumption}
Y(k) &= \sum_{m=1}^{M} Y_m(k)  + V(k), \\
Y_m(k) &= \sum_{k_1 + \hdots + k_m = k} H_m(\Omega_{k_1}, \hdots,\Omega_{k_m}) \prod_{i=1}^{m} U(k_i), 
\end{split}
\end{equation}
where $V(k)$ is complex circular output measurement noise with variance $\sigma^2_v$.
\end{assumption}

\begin{assumption}
For input and output spectra $Y(k)$ and $U(k)$ computed by an $N$-point DFT, the time domain input $u(t)$ is periodic in $N$, i.e.
$$u(t) = u(t+N) \; \; \; \forall t $$
This implies that the measured output will be transient-free, since the system is in steady-state with period length equal to the measurement window.
\end{assumption}

\begin{myDef}
For an $N$-point DFT, the set of DFT indices for which the input spectrum is non-zero is labelled $\mathbf{k_u} \subset \{1, \hdots, N/2M-1 \}$. The corresponding excited indices of the output spectrum are contained in $\mathbf{k_y}$.
\end{myDef}

\begin{remark}
The frequencies excited by the input signal, $u$, at which $H_m$ can be estimated, will be given by the set $\mathbf{\Omega} = \{ -\Omega_{\mathbf{k_u}} \; \; \Omega_{\mathbf{k_u}} \}$.
\end{remark}

\begin{myDef}
The $m$-dimensional tensor containing $H_m(\Omega_{k_1}, \hdots,\Omega_{k_m})$ $\forall \Omega_{k_1}, \hdots,\Omega_{k_m} \in \mathbf{\Omega}$ has a vectorized form which will be denoted by $H_m^{\mathcal{V}}$. The vectorization scheme remains arbitrary here, but will be discussed in Section \ref{MultidimensionalCov}.
\end{myDef}

\begin{assumption}
\label{GFRF_GaussianAssumption}
$H_m^{\mathcal{V}}$ is complex Gaussian distributed with zero mean and augmented covariance $\Sigma_m$, i.e.
\begin{equation} 
\label{GFRF_GaussianDefn}
H_m^{\mathcal{V}} \sim \mathcal{CN}(0,\Sigma_m) 
\end{equation}
\end{assumption}

\begin{assumption}
\label{GFRF_IndependenceAssumption}
Any two Gaussian GFRFs, $H_i^{\mathcal{V}}$ and $H_j^{\mathcal{V}}$, are independent for $i \neq j$.
\end{assumption}

\subsection{Designing Multidimensional Covariance Functions}

\label{MultidimensionalCov}

For multidimensional Volterra kernels, covariance functions have already been constructed in the time domain \cite{birpoutsoukis2017regularized}, by applying a diagonal/correlated (DC) structure \cite{chen2011kernel} along multiple perpendicular `regularizing directions.' The resulting covariance matrices are guaranteed to be valid and produce stable kernel realizations; two properties which we desire in the frequency domain context as well. These functions can be transformed to the frequency domain using a similar approach to that taken in \cite{lataire2016}, but the transformation will be more complex due to the required vectorization of time and frequency domain quantities in order to produce 2-dimensional covariance matrices. However, if the vectorization scheme is chosen carefully, the transformation matrix will possess a structure useful for software implementation. 

In this paper, for a quantity $X(i_1,\hdots,i_m)$ where $i_1,\hdots,i_m = 0,1,\hdots,N-1$, we define the vectorized quantity, $X^{\mathcal{V}}$, as the vector obtained by first incrementing $i_1$, then $i_2$, and so on. Now we are interested in finding the matrix, $F_m$, which represents the following transformation:
$$ H_m^{\mathcal{V}} = F_m h_m^{\mathcal{V}},$$
where $H_m$ is the multidimensional Fourier transform of Volterra kernel $h_m$. Under the considered vectorization scheme, $F_m$ can be constructed recursively from the $N \times N$ DFT matrix, which we denote $F_1$. The result is a matrix which retains the neat structure and symmetries of the 1-dimensional transform, as noted in \cite{briggs1995dft}. The recursive definition can be written as,
$$F_m = \begin{bmatrix} F_{m-1}(1,1) \cdot F_1 & \hdots &  F_{m-1}(1,N^{m-1}) \cdot  F_1  \\ \vdots & \ddots & \vdots \\ F_{m-1}(N^{m-1},1) \cdot F_1 & \hdots &  F_{m-1}(N^{m-1},N^{m-1})\cdot F_1 \end{bmatrix}, \; \; \; \; m = 2, 3, \hdots $$
where $F_{m}(i,j)$ indicates the $i,j$'th element of matrix $F_{m}$.

The \emph{vectorized} transformation can now be used to convert time domain covariance functions to the frequency domain as follows,
\begin{align}
 K_m &= \textbf{E} \{ H_m^{\mathcal{V}} {H_m^{\mathcal{V}}}^H\} = F_m \textbf{E} \{ h_m^{\mathcal{V}} {h_m^{\mathcal{V}}}^T \} F_m^H = F_m P_m F_m^H \\
C_m &= \textbf{E} \{ H_m^{\mathcal{V}}{H_m^{\mathcal{V}}}^T\} = F_m \textbf{E} \{ h_m^{\mathcal{V}} {h_m^{\mathcal{V}}}^T \} F_m^T = F_m P_m F_m 
\end{align}
where $P_m$ denotes the DC-based covariance structure of vectorized kernel $h_m^{\mathcal{V}}$ as designed in~\cite{birpoutsoukis2017regularized}.

\begin{remark}
In practice, symmetry must be enforced in both the time and frequency domain kernels to guarantee a unique Volterra series representation. This requires some modification of the transformation matrices, $F_m$, in order to consider only the \emph{unique} components of each kernel.
\end{remark}

\subsubsection{The Output Spectrum}

The output spectrum of $Y(\mathbf{k_y})$ can now be derived in a similar fashion to the linear case. However, we must first restructure the model equation, (\ref{ModelAssumption}), into a least squares framework, i.e.
$$ Y(\mathbf{k_y}) =  [\phi_1 \hdots \phi_M] [{H_1^{\mathcal{V}}}^T \hdots {H_M^{\mathcal{V}}}^T]^T + V(\mathbf{k_y})= \phi H + V,$$
where $\phi_m$ is an appropriate regressor containing the input spectrum products corresponding to $H_m^*$. Note that symmetry should also be enforced in the GFRFs here, which should be reflected in the design of the regressors. Extending this equation to the augmented output case gives,
\begin{equation}
\label{AugmentedModel}
\widetilde{Y}(\mathbf{k_y}) =  [\widetilde{\phi_1} \hdots \widetilde{\phi_M}] [\widetilde{H_1^{\mathcal{V}}}^T \hdots \widetilde{H_M^{\mathcal{V}}}^T]^T + \widetilde{V}(\mathbf{k_y}) = \widetilde{\phi}\widetilde{H} + \widetilde{V}
\end{equation}
where the augmented regressors are defined by
$$ \widetilde{\phi_m} = \begin{bmatrix} \phi_m & 0 \\ 0 & \overline{\phi_m} \end{bmatrix}.$$
From (\ref{AugmentedModel}), we can now derive the distribution of the output spectrum.

\begin{thm}
For a system given by (\ref{ModelAssumption}) with Gaussian $H_m^{\mathcal{V}}$ as described in (\ref{GFRF_GaussianDefn}), the output spectrum $Y(\mathbf{k_y})$ is complex normally distributed as,
\begin{equation}
\begin{split}
\label{OutputSpectrumThm}
Y(\mathbf{k_y}) &\sim \mathcal{CN}(0,\Sigma_Y), \\
\text{where } \Sigma_Y &= \widetilde{\phi}\Sigma_{tot} \widetilde{\phi}^H + \sigma^2_v I \\
\text{and } \Sigma_{tot} &= \begin{bmatrix} \Sigma_1 & &0 \\ & \ddots & \\ 0 & & \Sigma_M \end{bmatrix}
\end{split}
\end{equation}
\end{thm}
\begin{prf*}
Follows from the model equation in (\ref{AugmentedModel}), Assumptions \ref{GFRF_GaussianAssumption} and \ref{GFRF_IndependenceAssumption}, and the properties of complex normal distributions. \qed
\end{prf*}

\subsection{MAP Estimates of the GFRFs}

MAP estimates can be obtained from the joint distribution of $Y$ and $H$, by computing the mean of the conditional distribution $H|Y$. The result is provided in the following theorem.

\begin{thm}
The MAP estimate of $\widetilde{H}$ in (\ref{AugmentedModel}) is given by
\begin{equation}
\hat{{\widetilde{H}}}_{MAP} = \Sigma_{tot} \widetilde{\phi}^H \Sigma_Y^{-1} \widetilde{Y}
\end{equation}
\end{thm}
\begin{prf*}
Follows from the properties of joint complex distributions. \qed
\end{prf*}

\subsection{Hyperparameter Tuning using Marginal Likelihood}

Optimization of the hyperparameters describing each Gaussian process can be performed using the exact same methodology employed in \cite{lataire2016}, by maximizing the log marginal likelihood, i.e.
$$ \hat{\eta} = \text{arg } \underset{\eta}{\text{min }} \widetilde{Y}(\mathbf{k_y})^H \Sigma_Y^{-1}(\eta) \widetilde{Y}(\mathbf{k_y}) + \text{log det } \Sigma_Y(\eta). $$
The only notable difference is in the construction of $\Sigma_Y$, which is now formed using the complex regressor, $\widetilde{\phi}$, and a block-diagonal combination of each GFRFs covariance matrix.

It is important to note that the covariance matrices are first computed in the time domain as $P_m$, then transformed into the augmented frequency domain covariance, $\Sigma_m$, using transformation matrices $F_m$ as outlined in Section \ref{MultidimensionalCov}.

\section{Numerical Examples}

To demonstrate the estimation performance of Gaussian process regression for GFRFs, several systems with different nonlinear dynamics were simulated. The systems were all block-oriented in nature; Wiener, Hammerstein, and Wiener-Hammerstein, and the static nonlinear block had the form $f(x)=x^2$ in each case. This results in systems which contain only second order Volterra series terms.

\begin{figure}[hp]
\centering
\includegraphics[width=0.75\textwidth]{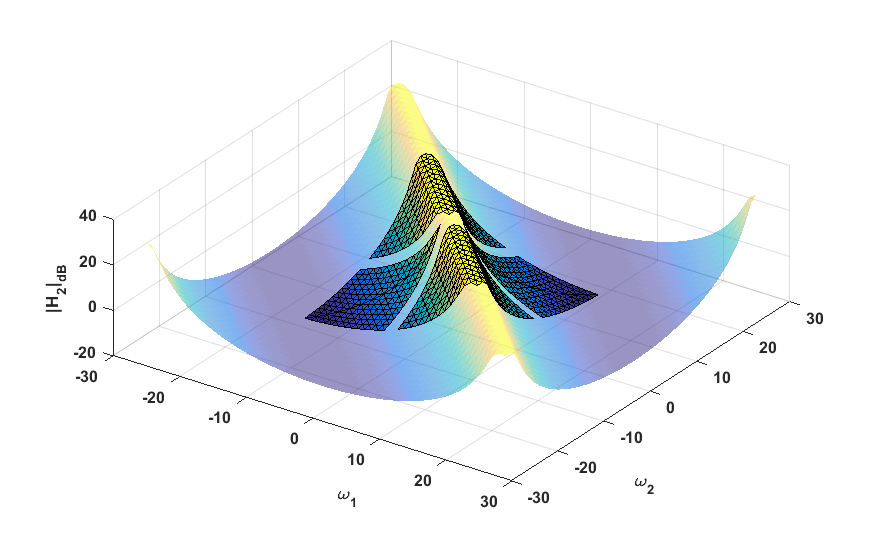}
\caption{True (transparent) and estimated GFRFs for a Hammerstein system}
\label{fig:HammersteinEstGFRF}
\end{figure}

\begin{figure}[hp]
\centering
\includegraphics[width=0.75\textwidth]{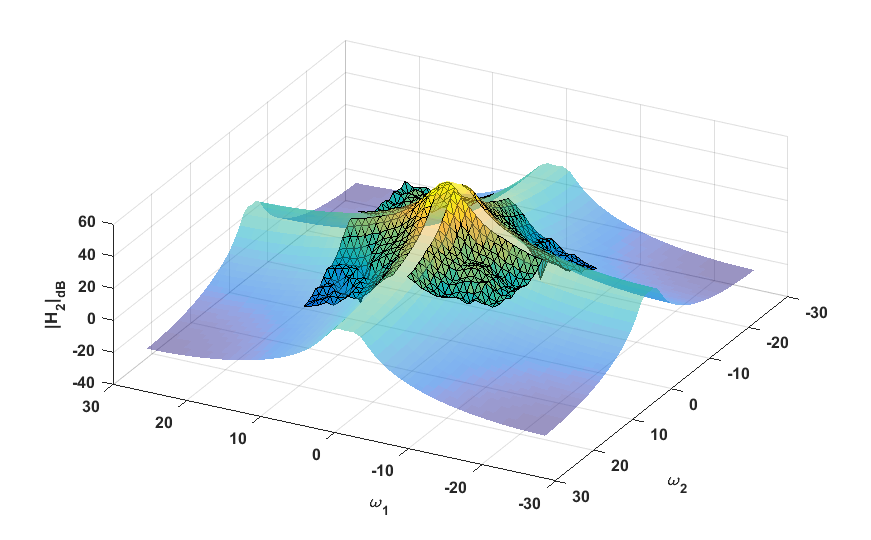}
\caption{True (transparent) and estimated GFRFs for a Wiener system}
\label{fig:WienerEstGFRF}
\end{figure}

\begin{figure}[hp]
\centering
\includegraphics[width=0.75\textwidth]{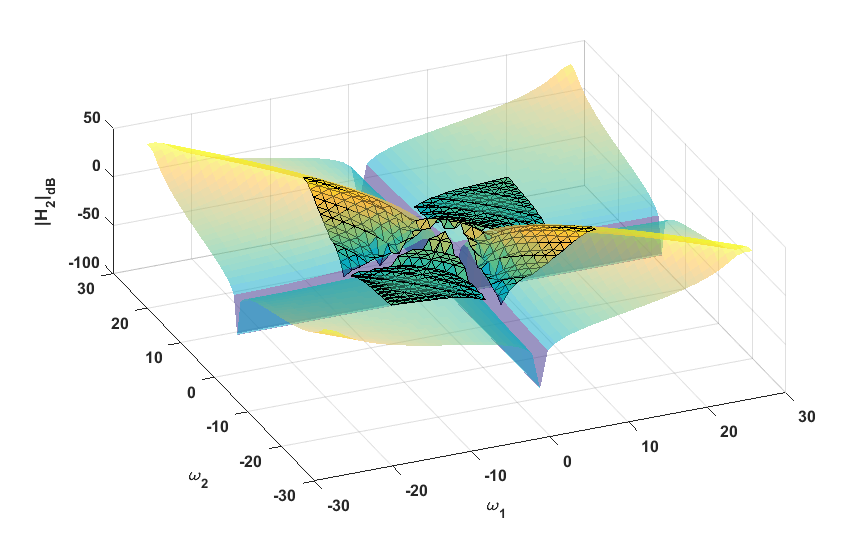}
\caption{True (transparent) and estimated GFRFs for a Wiener-Hammerstein system}
\label{fig:WHEstGFRF}
\end{figure}

The input was constructed as a random-phase multisine of length $N=55$ with 13 excited frequencies, allowing the estimation of $182$ unique GFRF parameters in each case. No measurement noise was added to the output spectrum, but Gaussian process regression is still required due to the rank deficiency of the estimation problem. For each nonlinear block structure, the resulting estimates have their magnitude plotted in Figures \ref{fig:HammersteinEstGFRF}, \ref{fig:WienerEstGFRF} and \ref{fig:WHEstGFRF} for a single input realization. The estimates are plotted on top of the true GFRF (transparent) for comparison, revealing a close match in each case. It is clear that tuning and transforming the time-domain covariance structures from \cite{birpoutsoukis2017regularized} leads to satisfactory GFRF estimation, even when the problem is severely rank deficient.

\section{Transients in a Second Order Volterra System} 

\begin{figure}[h]
\centering
\includegraphics[width=0.8\textwidth]{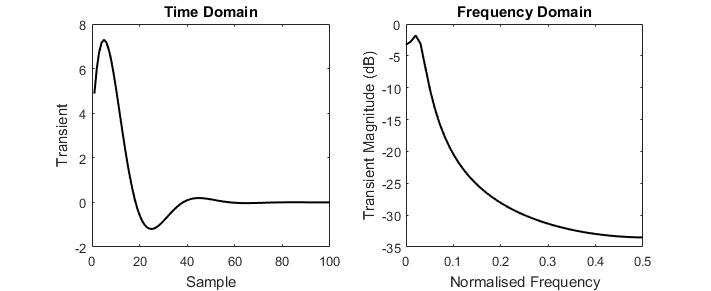}
\caption{Transient resulting from zero initial conditions in a linear filter}
\label{fig:LinearTrans}
\end{figure}

\begin{figure}[h]
\centering
\includegraphics[width=0.8\textwidth]{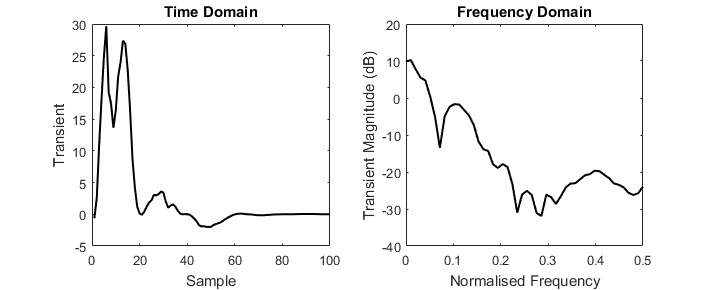}
\caption{Transient resulting from zero initial conditions in a second order Wiener system}
\label{fig:WienerTrans}
\end{figure}

While all previous results in this paper relied on the assumption of transient-free measurements from a periodic input excitation, one major benefit of Gaussian process regression for linear FRFs was the ability to estimate and remove transient functions for the non-periodic input case. However, this is only made possible by the approximation that the covariance of the transient function is a scaled version of the system covariance. While the approximation is well justified in the linear case, the same cannot be said in general for 2nd and higher order Volterra systems. The increased complexity of nonlinear transients can be observed in the examples of Figures \ref{fig:LinearTrans} and \ref{fig:WienerTrans}, which show transients resulting from a linear system and its second order Wiener counterpart.

In order to understand why this is the case, we first summarise the transient derivation for a linear system, and then derive an analytic expression for the transient response of a system whose output is described by a second order Volterra kernel. 

\subsection{Transient Expression for a Linear System}

The following is a summary of the derivation which can be found in \cite{lataire2016}. Consider a linear system described by its infinite impulse response, $h$, as,
$$ y(t) = \sum_{n=0}^{\infty} h(n) u(t-n),$$
where $u$ and $y$ are the input and output of the system respectively. It can be shown that the $N$-point DFT of $y(t), t = 0,1,\hdots,N-1$, which we will denote $Y(k)$, is related to the corresponding input DFT, $U(k)$, in the following way \cite{pintelon1997frequency}:
\begin{align*}
Y(k) &= \sum_{n=0}^{\infty} h(n) e^{\frac{-j 2 \pi k n}{N}} \sum_{t=0}^{N-1} u(t) e^{\frac{-j 2 \pi k t}{N}} + T(k) \\
&= H(k) U(k) +T(k),
\end{align*}
where $H(k) U(k)$ describes the steady state response to $u$, and $T(k)$ is a transient function given by (\cite{lataire2016}),
\begin{align}
T(k) &= \sum_{t=0}^{\infty} \underbrace{\sum_{n=t+1}^{\infty} h(n) f(t-n)}_{h^*(t)} e^{\frac{-j 2 \pi k t}{N}} \label{eq:LinearTransient} \\
\textrm{where } f(t) &= u(t) - u(t+N). \nonumber
\end{align} 

The term $h^*(t)$ is seen to be a free response of the system with impulse response $h_{n}$.

\subsection{Derivation of the Second Order Transient Expression}
\label{SecOrdTransDeriv}

Consider the noiseless output of a nonlinear system described by a second order Volterra kernel, i.e.
\begin{equation}
y(t) = \sum\limits_{\tau_{1}=0}^{\infty} \sum\limits_{\tau_{2}=0}^{\infty} h_{2}(\tau_{1},\tau_{2}) u(t - \tau_{1}) u(t - \tau_{2})
\end{equation}

Assume that $N$ samples of the input $u$ and output $y$ are available. Applying a DFT on the measured output signal $y(t), t = 0,1,\hdots,N-1$,  the spectral component of $y$ at DFT frequency $k$ is given by,
\begin{equation}
Y(k) = \sum\limits_{t=0}^{N-1} y(t) e^{-j \omega_{k} t} = \sum\limits_{t=0}^{N-1} (\sum\limits_{\tau_{1}=0}^{\infty} \sum\limits_{\tau_{2}=0}^{\infty} h_{2}(\tau_{1},\tau_{2}) u(t - \tau_{1}) u(t - \tau_{2})) e^{-j \omega_{k} t},
\label{eq:freq1}
\end{equation}
where $\omega_{k} = \frac{2 \pi k}{N}$. The term $e^{-j \omega_{k} t}$ can be split as follows:
\begin{equation}
\begin{aligned}
e^{-j \omega_{k} t} &= e^{-j \omega_{k_{1}} \tau_{1}} e^{-j \omega_{k-k_{1}} \tau_{2}} e^{-j \omega_{k_{1}} (t-\tau_{1})} e^{-j \omega_{k-k_{1}} (t-\tau_{2})} \\
& \textrm{with} \ \ \omega_{k_{1}} + \omega_{k-k_{1}} = \omega_{k}
\end{aligned}
\label{eq:freq2}
\end{equation}

Moreover, it holds that:

\begin{equation*}
\begin{aligned}
&\sum\limits_{t=0}^{N-1} u(t-\tau_{1}) e^{-j \omega_{k_{1}} t} = e^{-j \omega_{k_{1}} \tau_{1}} \bigg( \sum\limits_{t=0}^{N-1} u(t-\tau_{1}) e^{-j \omega_{k_{1}} (t - \tau_{1})} \bigg) \\
&= e^{-j \omega_{k_{1}} \tau_{1}} \bigg( \underbrace{\sum\limits_{t=0}^{N-1} u(t) e^{-j \omega_{k_{1}} t}}_{U(k_{1})} + \sum\limits_{t=-\tau_{1}}^{-1} (u(t)-u(t+N)) e^{-j \omega_{k_{1}} t} \bigg) \Leftrightarrow \\
& \underbrace{\sum\limits_{t=0}^{N-1} u(t-\tau_{1}) e^{-j \omega_{k_{1}} (t-\tau_{1})} }_{U_{t-\tau_{1}}(k_{1})} = U(k_{1}) + \sum\limits_{t=-\tau_{1}}^{-1} (u(t)-u(t+N)) e^{-j \omega_{k_{1}} t} \Leftrightarrow \\
& U_{t-\tau_{1}}(k_{1}) = U(k_{1}) + \sum\limits_{t=-\tau_{1}}^{-1} (u(t)-u(t+N)) e^{-j \omega_{k_{1}} t} \ \ \ \ \ \ \ \ \textrm{or}
\end{aligned}
\end{equation*}

\begin{equation}
\begin{aligned}
u(t-\tau_{1}) &= \sum\limits_{k_{1}=0}^{N-1} U_{t-\tau_{1}}(k_{1}) e^{j \omega_{k_{1}}(t-\tau_{1})} \Leftrightarrow \\
u(t-\tau_{1}) &= \sum\limits_{k_{1}=0}^{N-1} \bigg( U(k_{1}) + \sum\limits_{t=-\tau_{1}}^{-1} (u(t)-u(t+N)) e^{-j \omega_{k_{1}} t} \bigg) e^{j \omega_{k_{1}}(t-\tau_{1})}
\end{aligned}
\label{eq:freq3}
\end{equation}

Similarly it can be derived that:

\begin{equation}
\sum\limits_{t=0}^{N-1} u(t-\tau_{2}) e^{-j \omega_{k-k_{1}}(t-\tau_{2})} = U(k-k_{1}) + \sum\limits_{t=-\tau_{2}}^{-1} (u(t)-u(t+N)) e^{-j \omega_{k-k_{1}} t}
\label{eq:freq4}
\end{equation}

After substituting (\ref{eq:freq2}), (\ref{eq:freq3}) and (\ref{eq:freq4}) into (\ref{eq:freq1}) we obtain:

\begin{equation*}
\begin{aligned}
Y(k)&= \sum\limits_{k_{1}=0}^{N-1} \sum\limits_{\tau_{1}=0}^{\infty} \sum\limits_{\tau_{2}=0}^{\infty} h_{2}(\tau_{1},\tau_{2})  e^{-j \omega_{k_{1}} \tau_{1}} e^{-j \omega_{k-k_{1}} \tau_{2}} \\
&\ldots \bigg( U(k_{1}) + \sum\limits_{t^{'}=-\tau_{1}}^{-1} (\underbrace{u(t^{'})-u(t^{'}+N)}_{f(t^{'})}) e^{-j \omega_{k_{1}} t^{'}} \bigg)  \\
&\ldots \bigg( U(k-k_{1}) + \sum\limits_{t^{''}=-\tau_{2}}^{-1} (\underbrace{u(t^{''})-u(t^{''}+N)}_{f(t^{''})}) e^{-j \omega_{k-k_{1}} t^{''}} \bigg) \Leftrightarrow
\end{aligned}
\end{equation*}

\begin{equation}
\begin{aligned}
Y(k)&= \underbrace{ \sum\limits_{k_{1}=0}^{N-1} \bigg( \underbrace{\sum\limits_{\tau_{1}=0}^{\infty} \sum\limits_{\tau_{2}=0}^{\infty} h_{2}(\tau_{1},\tau_{2})  e^{-j \omega_{k_{1}} \tau_{1}} e^{-j \omega_{k-k_{1}} \tau_{2}}}_{H_{2}(k_{1},k-k_{1})} \bigg) U(k_{1}) U(k-k_{1}) }_{\textrm{Steady state response or} \ SS} \\
&+ \underbrace{ \sum\limits_{k_{1}=0}^{N-1} \sum\limits_{\tau_{1}=0}^{\infty} \sum\limits_{\tau_{2}=0}^{\infty} h_{2}(\tau_{1},\tau_{2})  e^{-j \omega_{k_{1}} \tau_{1}} e^{-j \omega_{k-k_{1}} \tau_{2}} U(k_{1}) \sum\limits_{t^{''}=-\tau_{2}}^{-1} f(t^{''}) e^{-j \omega_{k-k_{1}} t^{''}} }_{\textrm{Transient term 1 or} \ T_{1}} \\
&+ \underbrace{ \sum\limits_{k_{1}=0}^{N-1} \sum\limits_{\tau_{1}=0}^{\infty} \sum\limits_{\tau_{2}=0}^{\infty} h_{2}(\tau_{1},\tau_{2})  e^{-j \omega_{k_{1}} \tau_{1}} e^{-j \omega_{k-k_{1}} \tau_{2}} U(k-k_{1}) \sum\limits_{t^{'}=-\tau_{1}}^{-1} f(t^{'}) e^{-j \omega_{k_{1}} t^{'}} }_{\textrm{Transient term 2 or} \ T_{2}} \\
&+ \sum\limits_{k_{1}=0}^{N-1} \sum\limits_{\tau_{1}=0}^{\infty} \sum\limits_{\tau_{2}=0}^{\infty} h_{2}(\tau_{1},\tau_{2})  e^{-j \omega_{k_{1}} \tau_{1}} e^{-j \omega_{k-k_{1}} \tau_{2}} \times \\
& \times \underbrace{ \sum\limits_{t^{'}=-\tau_{1}}^{-1} f(t^{'}) e^{-j \omega_{k_{1}} t^{'}} \sum\limits_{t^{''}=-\tau_{2}}^{-1} f(t^{''}) e^{-j \omega_{k-k_{1}} t^{''}} }_{\textrm{Transient term 3 or} \ T_{3}}
\end{aligned}
\label{eq:freq6}
\end{equation}
which can be written compactly as:

\begin{equation}
\begin{aligned}
Y(k) &= SS(k) + \underbrace{T_{1}(k) + T_{2}(k) + T_{3}(k)}_{T(k)}\\
SS(k) &= \sum\limits_{k_{1}=0}^{N-1} H_{2}(k_{1},k-k_{1}) U(k_{1}) U(k-k_{1}) \\
T_{1}(k) &= \sum\limits_{k_{1}=0}^{N-1} \sum\limits_{\tau_{1}=0}^{\infty} \sum\limits_{\tau_{2}=0}^{\infty} h_{2}(\tau_{1},\tau_{2})  e^{-j \omega_{k_{1}} \tau_{1}} e^{-j \omega_{k-k_{1}} \tau_{2}} \sum\limits_{t^{'}= 0}^{N-1} u(t^{'}) e^{-j \omega_{k_{1}} t^{'}} \times \\
& \times \sum\limits_{t^{''}=-\tau_{2}}^{-1} f(t^{''}) e^{-j \omega_{k-k_{1}} t^{''}} \\
&= \sum\limits_{k_{1}=0}^{N-1} \sum\limits_{t^{'}= 0}^{\infty} \sum\limits_{t^{''}= 0}^{\infty} \underbrace{\Bigg[ \sum\limits_{\tau_{1}=t^{'}+1-N}^{t^{'}} \sum\limits_{\tau_{2}=t^{''}+1}^{\infty} h_{2}(\tau_{1},\tau_{2}) u(t^{'}-\tau_{1}) f(t^{''}-\tau_{2}) \Bigg]}_{h_1^*(t',t'')} \times \\
& \times e^{-j \omega_{k_{1}} t^{'}} e^{-j \omega_{k-k_{1}} t^{''}} \\
T_{2}(k) &= \sum\limits_{k_{1}=0}^{N-1} \sum\limits_{\tau_{1}=0}^{\infty} \sum\limits_{\tau_{2}=0}^{\infty} h_{2}(\tau_{1},\tau_{2})  e^{-j \omega_{k_{1}} \tau_{1}} e^{-j \omega_{k-k_{1}} \tau_{2}} \sum\limits_{t^{''}= 0}^{N-1} u(t^{''}) e^{-j \omega_{k-k_{1}} t^{''}} \times \\
& \times \sum\limits_{t^{'}=-\tau_{1}}^{-1} f(t^{'}) e^{-j \omega_{k_{1}} t^{'}} \\
&= \sum\limits_{k_{1}=0}^{N-1} \sum\limits_{t^{'}= 0}^{\infty} \sum\limits_{t^{''}= 0}^{\infty} \underbrace{\Bigg[ \sum\limits_{\tau_{1}=t^{'}+1}^{\infty} \sum\limits_{\tau_{2}=t^{''}+1-N}^{t^{''}} h_{2}(\tau_{1},\tau_{2}) f(t^{'}-\tau_{1}) u(t^{''}-\tau_{2}) \Bigg]}_{h_2^*(t',t'')} \times \\
& \times e^{-j \omega_{k_{1}} t^{'}} e^{-j \omega_{k-k_{1}} t^{''}} \\
T_{3}(k) &= \sum\limits_{k_{1}=0}^{N-1} \sum\limits_{t^{'}= 0}^{\infty} \sum\limits_{t^{''}= 0}^{\infty} \underbrace{\Bigg[ \sum\limits_{\tau_{1}=t^{'}+1}^{\infty} \sum\limits_{\tau_{2}=t^{''}+1}^{\infty} h_{2}(\tau_{1},\tau_{2}) f(t^{'}-\tau_{1}) f(t^{''}-\tau_{2}) \Bigg]}_{h_3^*(t',t'')} \times \\
& \times e^{-j \omega_{k_{1}} t^{'}} e^{-j \omega_{k-k_{1}} t^{''}}
\end{aligned}
\label{eq:freq7}
\end{equation}

\subsection{Equality of the transient terms $T_{1}$ and $T_{2}$}

Consider the term $T_{1}$ as written in (\ref{eq:freq6}), and apply the change of variables $k - k_{1} = k_{1}^{'} \Leftrightarrow k_{1} = k - k_{1}^{'}$. Moreover, interchange $\tau_{1}$ with $\tau_{2}$ due to the symmetry of $h_{2}(\tau_{1},\tau_{2}) = h_{2}(\tau_{2},\tau_{1}), \forall \tau_{1}, \tau_{2}$:
\begin{equation}
\begin{aligned}
T_{1}(k) &= \sum\limits_{k_{1}=0}^{N-1} \sum\limits_{\tau_{1}=0}^{\infty} \sum\limits_{\tau_{2}=0}^{\infty} h_{2}(\tau_{1},\tau_{2})  e^{-j \omega_{k_{1}} \tau_{2}} e^{-j \omega_{k-k_{1}} \tau_{1}} U(k_{1}) \sum\limits_{t^{''}=-\tau_{1}}^{-1} f(t^{''}) e^{-j \omega_{k-k_{1}} t^{''}} \\
&= \sum\limits_{k_{1}^{'}=k+1-N}^{k} \sum\limits_{\tau_{1}=0}^{\infty} \sum\limits_{\tau_{2}=0}^{\infty} h_{2}(\tau_{1},\tau_{2})  e^{-j \omega_{k - k_{1}^{'}} \tau_{2}} e^{-j \omega_{k_{1}^{'}} \tau_{1}} \sum\limits_{t^{''}=-\tau_{1}}^{-1} f(t^{''}) e^{-j \omega_{k_{1}^{'}} t^{''}} U(k - k_{1}^{'})\\
\end{aligned}
%\label{eq:freq8}
\end{equation}
Comparing the latter expression with the equivalent $T_{2}$ formulation,
\begin{equation}
\begin{aligned}
T_{2}(k) &= \sum\limits_{k_{1}=0}^{N-1} \sum\limits_{\tau_{1}=0}^{\infty} \sum\limits_{\tau_{2}=0}^{\infty} h_{2}(\tau_{1},\tau_{2}) e^{-j \omega_{k-k_{1}} \tau_{2}} e^{-j \omega_{k_{1}} \tau_{1}} \sum\limits_{t^{'}=-\tau_{1}}^{-1} f(t^{'}) e^{-j \omega_{k_{1}} t^{'}} U(k-k_{1}),
\end{aligned}
%\label{eq:freq9}
\end{equation}
we see that they differ only in the summation bounds of $k_1^{'}$. However, consider the following two facts:
\begin{itemize}
\item The DFT of the input signal $u(t)$ is periodic in the number of samples $N$, namely $U(k) = U(k+N)$. This implies that $\sum\limits_{k_{1} = 0}^{N-1} U(k-k_{1}) = \sum\limits_{k_{1} = M}^{M+N-1} U(k-k_{1}), \ \forall M$. By setting $M = k+1-N$ we obtain $\sum\limits_{k_{1} = 0}^{N-1} U(k-k_{1}) = \sum\limits_{k_{1} = k+1-N}^{k} U(k-k_{1})$.
\item The compex exponential function $e^{j \frac{2 \pi k}{N} t}$ is also periodic in $N$ since $e^{j \frac{2 \pi (k + z*N)}{N} t} = e^{j \frac{2 \pi k}{N} t}, \forall z \in \mathbb{Z}$
\end{itemize}
Under these conditions, it is clear that we can shift the summation bounds of $k_1^{'}$ without consequence, such that
\begin{equation}
\begin{aligned}
T_{1}(k) &= \sum\limits_{k_{1}^{'}=0}^{N-1} \sum\limits_{\tau_{1}=0}^{\infty} \sum\limits_{\tau_{2}=0}^{\infty} h_{2}(\tau_{1},\tau_{2})  e^{-j \omega_{k - k_{1}^{'}} \tau_{2}} e^{-j \omega_{k_{1}^{'}} \tau_{1}} \sum\limits_{t^{''}=-\tau_{1}}^{-1} f(t^{''}) e^{-j \omega_{k_{1}^{'}} t^{''}} U(k - k_{1}^{'}) \\
&= T_{2}(k)
\end{aligned}
\label{eq:freq10}
\end{equation}

\subsection{Qualitative Discussion of Transient Terms}

The analysis of the previous sections revealed that in the second order case, the formulation of the transient is significantly more complex than the linear case. In the frequency domain, the total transient can be viewed as the sum of three components, each of which comes from a two-dimensional time domain response that has been transformed and collapsed.

\begin{figure}[hp]
\centering
\includegraphics[width=0.9\textwidth]{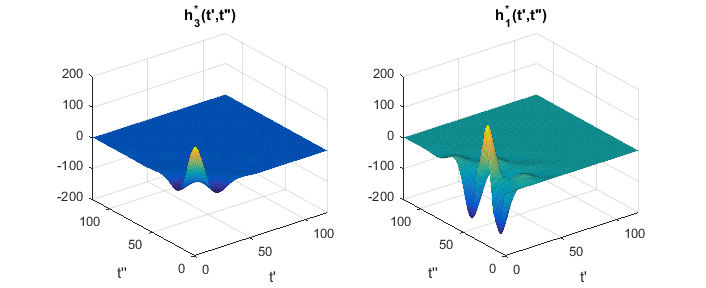}
\caption{The 2-dimensional time domain transient components for a Wiener system}
\label{fig:TDResponsesWiener}
\end{figure}

\begin{figure}[hp]
\centering
\includegraphics[width=0.75\textwidth]{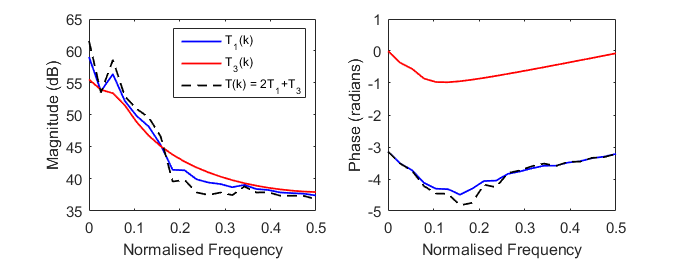}
\includegraphics[width=0.75\textwidth]{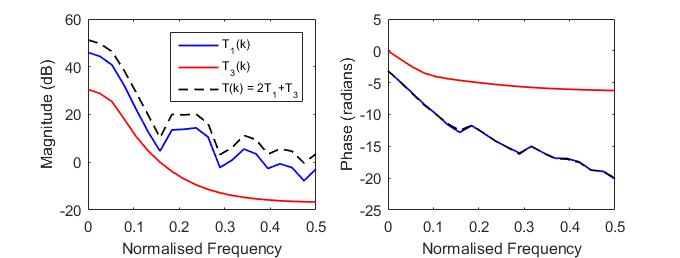}
\caption{Two realizations of the frequency domain transient components for a Wiener system}
\label{fig:FDResponsesWiener}
\end{figure}

The first two responses, $h^*_1$ and $h^*_2$, are asymmetric quantities which provide equal contributions to the transient, and depend directly on the input inside the measured window. The third response, $h_3^*$, represents a natural extension of the linear transient expression in (\ref{eq:LinearTransient}), and consequently maintains the desirable property of having similar structure to the underlying system response, $h_2(\tau_1,\tau_2)$. Consequently, the corresponding output transient, $T_3(k)$, will also share the smoothness properties of linear transients. Example responses are visible in Figure \ref{fig:TDResponsesWiener} for a second order Wiener system, showing the contrast in complexity between $h_3^*$ and the assymetric $h_1^*$. The resulting frequency domain functions, $T_1(k)$, $T_3(k)$ and $T(k)$, are plotted in Figure \ref{fig:FDResponsesWiener} for two Gaussian input realizations.

In general, the existence of the $h^*_1$ and $h^*_2$ terms will complicate the form of the transient in time and frequency domains, which in turn necessitates more sophisticated algorithms for detecting and removing the transient function. As the nonlinear order is increased further, so will the complexity of the transients generated from non-periodic excitation.

\subsection{Special case of diagonal Volterra kernels: the Hammerstein block structure}

In the special case where the underlying dynamics can be described by the Hammerstein block structure, it can be shown that the spectrum of the system output is given by:

\begin{equation*}
\begin{aligned}
Y(k) &= SS(k) - T_{3}(k) + 2 R_{H}(k) \\
R_{H}(k) &= 
\begin{cases}
0 , \ \textrm{when} \ u(t) = 0, \forall t<0\\ 
\sum\limits_{\tau=0}^{\infty} h_{2}(\tau,\tau) e^{-j \omega_{k} \tau} \sum\limits_{k_{1}=0}^{N-1} \sum\limits_{t^{'}=-\tau}^{-1} u(t^{'}) e^{-j \omega_{k_{1}} t^{'}} \sum\limits_{t^{''}=-\tau}^{-1} f(t^{''}) e^{-j \omega_{k-k_{1}} t^{''}}, \ \textrm{otherwise}
\end{cases}
\end{aligned}
\end{equation*}

\begin{prf*}

From (\ref{eq:freq6}) we get:
\begin{equation}
\begin{aligned}
T_{3}(k) &= \sum\limits_{k_{1}=0}^{N-1} \sum\limits_{\tau_{1}=0}^{\infty} \sum\limits_{\tau_{2}=0}^{\infty} h_{2}(\tau_{1},\tau_{2})  e^{-j \omega_{k_{1}} \tau_{1}} e^{-j \omega_{k-k_{1}} \tau_{2}} \underbrace{ \sum\limits_{t^{'}=-\tau_{1}}^{-1} f(t^{'}) e^{-j \omega_{k_{1}} t^{'}}}_{f_{\tau_{1}}} \times \\
& \times \sum\limits_{t^{''}=-\tau_{2}}^{-1} f(t^{''}) e^{-j \omega_{k-k_{1}} t^{''}}
\end{aligned}
\label{eq:freq8}
\end{equation}
with
\begin{equation}
\begin{aligned}
f_{\tau_{1}} &= \sum\limits_{t^{'}=-\tau_{1}}^{-1} f(t^{'}) e^{-j \omega_{k_{1}} t^{'}} = \sum\limits_{t^{'}=-\tau_{1}}^{-1} (u(t^{'}) - u(t^{'} + N)) e^{-j \omega_{k_{1}} t^{'}} \\
&= \sum\limits_{t^{'}=-\tau_{1}}^{-1} u(t^{'}) e^{-j \omega_{k_{1}} t^{'}} - \sum\limits_{t^{'}=-\tau_{1}}^{-1} u(t^{'}+N) e^{-j \omega_{k_{1}} t^{'}} \\
&= \sum\limits_{t^{'}=-\tau_{1}}^{-1} u(t^{'}) e^{-j \omega_{k_{1}} t^{'}} - \sum\limits_{t^{'}=N-\tau_{1}}^{N-1} u(t^{'}) e^{-j \omega_{k_{1}} t^{'}} \ (\textrm{periodicity of} \ \ e^{-j \omega_{k_{1}} t^{'}} \ \textrm{in} \ N) \\
&= \sum\limits_{t^{'}=-\tau_{1}}^{-1} u(t^{'}) e^{-j \omega_{k_{1}} t^{'}} - \underbrace{\sum\limits_{t^{'}= 0}^{N-1} u(t^{'}) e^{-j \omega_{k_{1}} t^{'}}}_{U(k_{1})}  + \sum\limits_{t^{'}= 0}^{N-\tau_{1}-1} u(t^{'}) e^{-j \omega_{k_{1}} t^{'}} \Leftrightarrow \\
f_{\tau_{1}} &= -U(k_{1}) + \sum\limits_{t^{'}= 0}^{N-\tau_{1}-1} u(t^{'}) e^{-j \omega_{k_{1}} t^{'}} + \sum\limits_{t^{'}=-\tau_{1}}^{-1} u(t^{'}) e^{-j \omega_{k_{1}} t^{'}}
\end{aligned}
\label{eq:freq9}
\end{equation}

Substituting (\ref{eq:freq9}) into (\ref{eq:freq8}) leads to:
\begin{equation*}
\begin{aligned}
T_{3}(k) &= - T_{1}(k) + Q(k) + R(k) \\
Q(k) &= \sum\limits_{k_{1}=0}^{N-1} \sum\limits_{\tau_{1}=0}^{\infty} \sum\limits_{\tau_{2}=0}^{\infty} h_{2}(\tau_{1},\tau_{2})  e^{-j \omega_{k_{1}} \tau_{1}} e^{-j \omega_{k-k_{1}} \tau_{2}} \sum\limits_{t^{'}= 0}^{N-\tau_{1}-1} u(t^{'}) e^{-j \omega_{k_{1}} t^{'}} \times \\
& \times \sum\limits_{t^{''}=-\tau_{2}}^{-1} f(t^{''}) e^{-j \omega_{k-k_{1}} t^{''}}
\end{aligned}
\end{equation*}

\begin{equation*}
\begin{aligned}
R(k) &= \sum\limits_{k_{1}=0}^{N-1} \sum\limits_{\tau_{1}=0}^{\infty} \sum\limits_{\tau_{2}=0}^{\infty} h_{2}(\tau_{1},\tau_{2})  e^{-j \omega_{k_{1}} \tau_{1}} e^{-j \omega_{k-k_{1}} \tau_{2}}  \sum\limits_{t^{'}=-\tau_{1}}^{-1} u(t^{'}) e^{-j \omega_{k_{1}} t^{'}} \times \\
& \times \sum\limits_{t^{''}=-\tau_{2}}^{-1} f(t^{''}) e^{-j \omega_{k-k_{1}} t^{''}}
\end{aligned}
\end{equation*}

with $T_1(k)$ defined in (\ref{eq:freq6}). This equation holds for any second order Volterra kernel. In the case of a Hammerstein system, we know that $h_{2}(\tau_{1},\tau_{2}) = 0 \ \ \forall \tau_{1} \neq \tau_{2}$. Hence we are interested only in the contribution of the diagonal elements $h_{2}(\tau,\tau)$ to the summations. Under these conditions, $Q(k)$ can be reduced as follows,
\begin{equation*}
\begin{aligned}
Q_{H}(k) &= \sum\limits_{k_{1}=0}^{N-1} \sum\limits_{\tau=0}^{\infty} h_{2}(\tau,\tau)  e^{-j \omega_{k_{1}} \tau} e^{-j \omega_{k-k_{1}} \tau} \sum\limits_{t^{'}= 0}^{N-\tau-1} u(t^{'}) e^{-j \omega_{k_{1}} t^{'}} \times \\
& \times \sum\limits_{t^{''}= N-\tau}^{N-1} f(t^{''}-N) e^{-j \omega_{k-k_{1}} t^{''}} \\
&= \sum\limits_{\tau=0}^{\infty} h_{2}(\tau,\tau)  e^{-j \omega_{k} \tau} \sum\limits_{t^{'}= 0}^{N-\tau-1} \sum\limits_{t^{''}= N-\tau}^{N-1} u(t^{'}) f(t^{''}-N) e^{-j \omega_{k} t^{''}}  \sum\limits_{k_{1}=0}^{N-1} e^{-j \omega_{k_{1}}(t^{''} - t^{'})} \\
&= 0,
\end{aligned}
\end{equation*}
since (observing that $(t^{''} - t^{'}) \in \mathbb{Z}$ and $0 < t^{''} - t^{'} < N$ due to the bounds of the summations),
\begin{equation*}
\begin{aligned}
\sum\limits_{k_{1}=0}^{N-1} e^{-j \omega_{k_{1}}(t^{''} - t^{'})} &= \sum\limits_{k_{1}=0}^{N-1} \bigg(e^{-j \frac{2 \pi (t^{''} - t^{'})}{N}} \bigg)^{k_{1}}  = \frac{1 - \bigg(e^{-j \frac{2 \pi (t^{''} - t^{'})}{N}} \bigg)^{N}}{1 - e^{-j \frac{2 \pi (t^{''} - t^{'})}{N}}} \\
&= \frac{1 - e^{-j 2 \pi (t^{''} - t^{'})}} {1 - e^{-j \frac{2 \pi (t^{''} - t^{'})}{N}}} = \frac{1 - 1}{1 - \underbrace{e^{-j \frac{2 \pi (t^{''} - t^{'})}{N}}}_{\neq 1}} = 0.
\end{aligned}
\end{equation*}

The term $R(k)$ can be similarly elaborated to show that, for the Hammerstein case,
\begin{flalign*}
&& R_{H}(k) &= \sum\limits_{k_{1}=0}^{N-1} \sum\limits_{\tau=0}^{\infty} h_{2}(\tau,\tau) e^{-j \omega_{k} \tau} \sum\limits_{t^{'}=-\tau}^{-1} u(t^{'}) e^{-j \omega_{k_{1}} t^{'}} \sum\limits_{t^{''}=-\tau}^{-1} f(t^{''}) e^{-j \omega_{k-k_{1}} t^{''}} && \\
&& &= 0 \textrm{ when } u(t) = 0 \ \forall t<0. &&
\end{flalign*}

Now, the output spectrum is the summation of the steady state reponse with all transient components, i.e.
\begin{flalign*}
&& Y(k) &= SS(k) + T_{1}(k) + T_2(k) + T_3(k) && \\
&&  &= SS(k) + 2 T_{1}(k) + T_3(k) && \\
&&  &= SS(k) + 2(-T_{3}(k)+Q_H(k)+R_H(k)) + T_3(k) && \\
&& &=  SS(k) - T_3(k) + 2 R_H(k) &&\qed
\end{flalign*}

\end{prf*}

\subsubsection{Qualitative Discussion of the Hammerstein Case}
\label{subsubsec:Qualitative_Discussion}

Observing the two possible transient expressions for the Hammerstein case, it is clear that the total transient will always be similar to the linear case, since $T_3(k)$ was already seen to be a `free response' to initial conditions, and $R_H(k)$ will either be $0$ or another similar free response. This result is consistent with the intuition which can be obtained from a block-oriented perspective. The static nonlinearity performs a nonlinear scaling on the input (and initial conditions), but the output is still generated by a linear filter applied to a signal, and so the transient, accordingly, will look `linear'.

\begin{figure}[hp]
\centering
\includegraphics[width=0.8\textwidth]{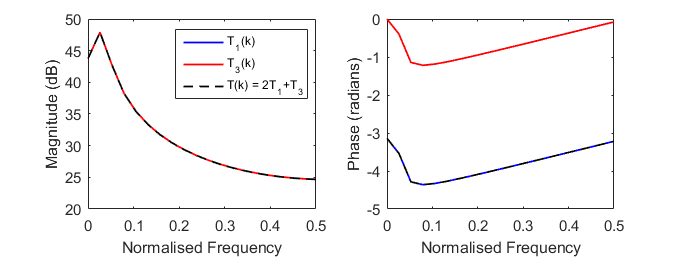}
\includegraphics[width=0.8\textwidth]{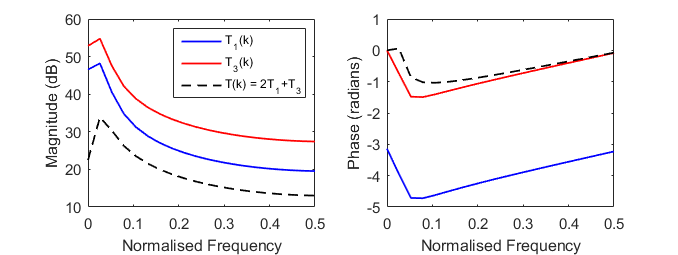}
\caption{Hammerstein transients in the frequency domain for zero (top) and non-zero (bottom) initial conditions}
\label{fig:FDResponsesHammerstein}
\end{figure}

Some Hammerstein transients are plotted in Figure \ref{fig:FDResponsesHammerstein} for both zero and non-zero initial conditions. In the former case, the total transient has equal magnitude with $T_1$ and $T_3$ (since $R_H(k) = 0$), and in the latter case, the total transient is another smooth function of frequency. 

%\bibliographystyle{ieeetr}        % Include this if you use bibtex 
%\newpage
%\bibliography{Blah} 

\bibliographystyle{plain}
\bibliography{Automatica2015}

\end{document}